\documentclass[twocolumn,prstab,floatfix]{revtex4}
\usepackage{graphicx}
\usepackage{amsmath}
\usepackage{latexsym}
\usepackage{amsfonts}
\usepackage{amssymb}

\begin{document}

\title{Strength of Higher-Order Spin-Orbit Resonances}
\author{Georg H.~Hoffstaetter} 
\affiliation{Department of Physics, Cornell University, Ithaca/NY}
\author{Mathias Vogt} 
\affiliation{DESY, Hamburg/FRG}

\begin{abstract}
When polarized particles are accelerated in a synchrotron, the spin
precession can be periodically driven by Fourier components of the
electromagnetic fields through which the particles travel. This leads
to resonant perturbations when the spin-precession frequency is close
to a linear combination of the orbital frequencies.  When such
resonance conditions are crossed, partial depolarization or spin flip
can occur. The amount of polarization that survives after resonance
crossing is a function of the resonance strength and the crossing
speed. This function is commonly called the Froissart-Stora
formula. It is very useful for predicting the amount of polarization
after an acceleration cycle of a synchrotron or for computing the
required speed of the acceleration cycle to maintain a required amount
of polarization.  However, the resonance strength could in general
only be computed for first-order resonances and for synchrotron
sidebands. When Siberian Snakes adjust the spin tune to be
$\frac{1}{2}$, as is required for high energy accelerators,
first-order resonances do not appear and higher-order resonances
become dominant.  Here we will introduce the strength of a
higher-order spin-orbit resonance, and also present an efficient
method of computing it. Several tracking examples will show that the
so computed resonance strength can indeed be used in the
Froissart-Stora formula. HERA-p is used for these examples which
demonstrate that our results are very relevant for existing
accelerators.
\end{abstract}

\maketitle

\section{Introduction}

In this paper we want to introduce the strength of higher-order spin
orbit resonances which we want to use in the Froissart-Stora formula
to compute how much polarization is lost when a resonance is
crossed. For first-order resonances the definition and computation of
the resonance strength is relatively simple
\cite{courant80,hoffstaetter99i}, for higher-order resonances it is
much more elaborate. We will need to use the invariant spin field,
also called the $\vec n$-axis \cite{derbenev72}, the amplitude
dependent spin tune and the periodic coordinate system over phase
space that determines the spin tune \cite{yokoya86a}. These concepts
are therefore quickly reviewed in this introduction.

While a polarized particle moves along the azimuth $\theta=\frac{2\pi
l}{L}$ of the storage ring's closed orbit with path length $l$ and
total length $L$, its semi-classical spin precesses according to the
T-BMT equation \cite{thomas27,bargmann59}
\begin{equation}
\frac{\text{d}}{\text{d}\theta}\vec S = \vec\Omega_0(\theta)\times\vec S\ .
\end{equation}
The spin direction that is periodic after one turn is referred to as
$\vec n_0(\theta)$.  If the spin has any other direction, it precesses
around $\vec n_0$. The numbers of precessions that occur during one
turn is referred to as the closed orbit spin tune $\nu_0$. To describe
the precession, a right handed system of orthogonal unit vectors
$(\vec m,\vec l,\vec n_0)$ is introduced for any azimuth.  The two
vectors $\vec l(\theta)$ and $\vec m(\theta)$ precess around $\vec
n_0$ according to the T-BMT equation so that they would have rotated
$\nu_0$ times after one turn. However a precession is added that
continuously winds back $\nu_0$ precessions. These vectors are
therefore periodic in the azimuth and
\begin{eqnarray}
\frac{\text{d}}{\text{d}\theta}\vec m
&=&
(\vec\Omega_0-\nu_0\vec n_0)\times\vec m\ ,\\
\frac{\text{d}}{\text{d}\theta}\vec l
&=&
(\vec\Omega_0-\nu_0\vec n_0)\times\vec l\ .
\end{eqnarray}

Since particles on the closed orbit have spins that precess around
$\vec n_0$, the product $s_3=\vec S(\theta)\cdot\vec n_0(\theta)$ is
an invariant, i.e.~does not depend on $\theta$. It can be shown that
it is also an adiabatic invariant
\cite{hoffstaetter00,hoffstaetter02a,neistadt75}, i.e.~it
hardly changes when parameters of the system, like the storage energy,
are slowly changed.

This concept of an invariant spin direction, a spin tune, a periodic
system of unit vectors and an adiabatic invariant can be extended to
particles that do not move on the closed orbit but oscillate around
this orbit and whose motion is thus described by phase space
trajectories $\vec z(\theta)$. The T-BMT equation for spin motion then
depends on the phase space trajectory
\begin{equation}
\frac{\text{d}}{\text{d}\theta}\vec S = \vec\Omega(\vec
z,\theta)\times\vec S\ .
\end{equation}

If the vector field $\vec f(\vec z,\theta)$ with $|\vec f|=1$
describes the spin distribution in a particle beam, it is called a
spin field and satisfies the T-BMT equation
\begin{equation}
\frac{\text{d}}{\text{d}\theta}\vec f(\vec z(\theta),\theta) =
\vec\Omega(\vec z,\theta)\times\vec f\ .
\end{equation}
A special spin field that is periodic from turn to turn is called the
invariant spin field $\vec n$,
\begin{equation}
\vec n(\vec z,\theta+2\pi) = \vec n(\vec z,\theta)\ .
\end{equation}

Particles that travel along the trajectory $\vec z(\theta)$ have spins
that precess around $\vec n(\vec z(\theta),\theta)$. Describing this
precession and even the number of precessions in one turn starting at
$\vec z(\theta_0)$ is not trivial, since the particle has a new phase
space point $\vec z(\theta_0+2\pi)$ after one turn. An orthogonal set
of unit vectors $(\vec u_1,\vec u_2,\vec n)$ has to be defined for
each phase space point and for each azimuth to determine spin
precession angles.

If the unit vectors $\vec u_1$ and $\vec u_2$ would satisfy the T-BMT
equation along each phase space trajectory starting at $\vec z_i$ and
ending at $\vec z_f$ after one turn, these vectors would precess
around $\vec n$ and after one turn $\vec u_i(\vec
z_f,\theta_0+2\pi)$ would have some angle $2\pi\tilde\nu(\vec
z_0)$ with respect to the initial unit vectors $\vec u_i(\vec
z_f,\theta_0)$ at the same phase space point.

The rotation angle $2\pi\tilde\nu$ is not well defined, since the
direction of the $\vec u_i$ before and after the turn is only required
to be perpendicular to $\vec n$, but has a free angular orientation in
the orthogonal plain.  This free orientation for each phase space
point can (under certain general conditions
\cite{hoffstaetter00,vogt00,BEH04}) be chosen to make the number of
rotations $\nu$ independent of the orbital phase variables
$\vec\Phi$. It then only depends on the amplitudes $\vec J$ of the orbital
motion and is therefore called the amplitude dependent spin tune
$\nu(\vec J)$.

To obtain a periodic set of unit vectors, the described precession of
the unit vectors is again augmented by continuously winding back $\nu$
spin precessions during one turn,
\begin{equation}
\frac{\text{d}}{\text{d}\theta}\vec u_i(\vec z(\theta),\theta) =
[\vec\Omega(\vec z,\theta)-\nu(\vec J)\vec n(\vec z,\theta)]\times\vec
u_i\ .
\end{equation}

Since spins precess around $\vec n$, the product \mbox{$J_S=\vec
S(\theta)\cdot\vec n(\vec z(\theta),\theta)$} is an invariant of
motion, i.e.~it does not change with $\theta$. It can be shown,
however, that it is also an adiabatic invariant
\cite{hoffstaetter00,hoffstaetter02a,neistadt76}, i.e.~it hardly
changes when system parameters like the storage energy change
sufficiently slowly.  This has strong implications. When a beam is
polarized parallel to the invariant spin field $\vec n(\vec z,E_i)$ at
some initial energy $E_i$ and the storage energy is increased slowly,
the beam will be polarized parallel to $\vec n(\vec z,E_f)$ at the
final energy $E_f$.

This is a very important property since a beam in such a polarization
state will have the average polarization $P_{\mathrm lim}=<\vec n>$ after
acceleration, which can be large even if this average polarization is
small at intermediate energies.

\section{The Single Resonance Model (SRM)}
\label{sc:srmodel}

\subsection{Fourier Expansion of Spin Perturbations}

The quantities $\vec n$, $\nu$, $\vec u_1$, $\vec u_2$ and $J_S$ will
be computed for an analytically solvable model and the adiabatic
invariance will be illustrated by letting a parameter of this model
change. Since this model leads to the Froissart-Stora formula, a
comparison of its equations with the equations of general spin
dynamics leads to the introduction of higher-order resonance strengths
that can be used in the Froissart-Stora formula.

The spin precession vector for particles which oscillate around the
closed orbit can be decomposed into the closed-orbit contribution
$\vec\Omega_0$ and a part $\vec\omega$ due to the particles'
oscillations, $\vec\Omega(\vec z,\theta) = \vec\Omega_0(\theta) +
\vec\omega(\vec z,\theta)$.  In the $(\vec m,\vec l,\vec n_0)$ system
we write
\begin{equation}
\vec S=s_1\vec m+s_2\vec l+s_3\vec n_0\; , \ \
\vec\omega=\omega_1\vec m+\omega_2\vec l+\omega_3\vec n_0\ .
\label{eq:smln0}
\end{equation}
With the complex notation $\hat s=s_1+\text{i} s_2$ and
$\omega=\omega_1+\text{i} \omega_2$, the equation of spin motion is
\begin{equation}
\vec\Omega\times\vec S
=
\vec m\frac{\text{d}}{\text{d}\theta}s_1+
\vec l\frac{\text{d}}{\text{d}\theta}s_2+
\vec n_0\frac{\text{d}}{\text{d}\theta}s_3+
(\vec\Omega_0-\nu_0\vec n_0)\times\vec S
\end{equation}
and the equation of motion for $\hat s$ is obtained by multiplication
with $\vec m+\text{i} \vec l$, and taking into account that $s_3=\sqrt{1-|\hat
  s|^2}$,
\begin{equation}
\frac{\text{d}}{\text{d}\theta}\hat s=i(\nu_0+\omega_3)\hat s-\text{i} \omega\sqrt{1-|\hat s|^2}\; .
\label{eq:hatbmt}
\end{equation}
In a coordinate system that rotates by $\nu_0\theta$, this equation
becomes
\begin{equation}
\hat s_0=\text{e}^{-\text{i} \nu_0\theta}\hat s\; ,\ \
\frac{\text{d}}{\text{d}\theta}\hat s_0=i\omega_3\hat s_0
                         -\text{i} \text{e}^{-\text{i} \nu_0\theta}
\omega\sqrt{1-|\hat s_0|^2}\; .
\end{equation}
Spin precession on the closed orbit ($\omega=0$) leads to a constant
$\hat s_0$ due to the left equation. The right equation describes
additional precessions due to phase space motion.

If the motion in phase space can be transformed to action-angle
variables, the spin precession vector $\vec\omega(\vec
J,\vec\Phi,\theta)$ for particles which oscillate around the closed
orbit is a $2\pi$-periodic function of $\vec\Phi$ and $\theta$.  The
Fourier spectrum of $\omega(\vec J,\Phi_0+\vec Q\theta,\theta)$ has
frequencies $\kappa=j_0+\vec j\cdot\vec Q$ where the $j_k$ are integers and
$\vec Q$ describes the tunes of synchrotron and betatron oscillations.
The integer contributions $j_0$ are due to the $2\pi$ periodicity of
$\vec\omega$ in $\theta$ and give rise to so-called imperfection
resonances. The contributions $\vec j\cdot\vec Q$ of integer multiples
of the orbit tunes are due to the $2\pi$ periodicity of $\vec\omega$
in the orbital phases $\Phi_k$ and give rise to so-called intrinsic
resonances \cite{courant80}.  When one of the Fourier frequencies is
nearly in resonance with $\nu_0$, one component of
$\text{e}^{-\text{i} \nu_0\theta}\omega$ is nearly constant. Then it
can be a good approximation to drop all other Fourier components since
their influence on spin motion can average to zero so that they are in
effect less dominant.  This is referred to as the single resonance
approximation.  Note that this approximation can only be good when the
domains of influence of individual resonances are well separated.
This model corresponds to the rotating field approximation often used
to discuss spin resonance in solid state physics \cite{abragam61}.
Note also that for a conventional flat ring, the first-order
resonances due to vertical motion dominate and therefore the Fourier
components with frequencies $\kappa=j_0\pm Q_y$ are often of most
interest.

The amplitude of a single Fourier contribution is sometimes called the
resonance strength. This is misleading since generally it cannot be
used in the Froissart-Stora formula. The fact that the Fourier
component is not the resonance strength manifest itself clearly in
models where where $\vec\omega$ is linear and has only first-order
Fourier components, i.e.~those with $\sum_{k=1}^3|j_k|=1$.  Such a
$\vec\omega$ can lead to depolarization or spin flip at first-order
resonances but also at higher-order resonances
\cite{mane87b,lee85,lee93a,lee97a}. The strength of these resonances
that might be use in the Froissart-Stora formula can clearly not be
determined by the higher order Fourier coefficients, i.e.~those where
$\sum_{k=1}^3|j_k|>1$, since those are zero.  In fact all examples of
higher-order resonances that will be shown in this paper were computed
for such a linear model of HERA-p with Siberian Snakes
\cite{future_epac00}.

A higher order resonance can thus be created either by a higher-order
Fourier component or by feed-up of lower order components. Such a
feed-up can occur due to the inherent non-commutativity of three
dimensional rotations or equivalently due to the nonlinearity of the
mapping from the unit sphere to the complex plane which gives rise to
the square root term in the equation of motion
(\ref{eq:hatbmt}). Obtaining a resonance strength $\epsilon_\kappa$
that can be used to describe depolarization therefore has to include
all these feed-up effects. Before the following investigations it was
not clear whether a Froissart-Stora formula with some resonance
strength $\epsilon_\kappa$ could be applied to crossing such
higher-order resonances.  But even if it can be applied, it is clear
that the resonance strength cannot be obtained from a Fourier
coefficient of $\vec\omega$ in (\ref{eq:hatbmt}).  Moreover, in high
energy accelerators, the $n$-th order Fourier coefficients of
$\vec\omega$ are not even the dominant contribution to the strength of
a $n$-th order resonance.  Usually the former contain $G\gamma$,
whereas the feed-up contributions from combining $m$ lower order
harmonics, contain $(G\gamma)^m$, which can be an exceedingly large
number.

Only for first-order resonances, where $\sum_{k=1}^3|j_k|=1$, there is
no feed-up contribution and the Fourier components can generally be
used in the Froissart-Stora formula and there are different straight
forward ways of computing $\epsilon_\kappa$ in that case
\cite{courant80,hoffstaetter99i}.

\subsection{Solutions for the SRM}

The analytically solvable model advertised above is usually called the
single resonance model (SRM).  It has $\vec\Omega_0=\nu_0\vec n_0$ and
an $\vec\omega$ which only has one Fourier contribution,
$\vec\omega=\epsilon_\kappa(\vec m\cos\Phi +\vec l\sin\Phi)$, with
$\Phi=j_0\theta+\vec j\cdot\vec\Phi+\Phi_0$.  Note that the modulus of
its higher-order Fourier coefficient is denoted as $\epsilon_\kappa$
since there are no lower order coefficients that could contribute to
the resonance strength by a feed-up precess. Any dependence on the
orbital actions can be expressed by $\epsilon_\kappa(\vec J)$.

This $\vec\omega$ is perpendicular to $\vec n_0$ and tilts spins away
from $\vec n_0$. Since $\frac{\text{d}}{\text{d}\theta}\vec\Phi=\vec
Q$, the frequency is $\kappa=j_0+\vec j\cdot\vec Q$ and the equation
of motion (\ref{eq:hatbmt}) becomes
\begin{equation}
\frac{\text{d}}{\text{d}\theta}\hat s=i\nu_0\hat s-\text{i} \epsilon_\kappa
\text{e}^{\text{i} (\kappa\theta+\Phi_0)}\sqrt{1-|\hat s|^2}\; .
\label{eq:hatsrm}
\end{equation}
When the coordinates in the $[\vec
m,\vec l,\vec n_0]$ system are arranged in column vectors
\cite{mane88, hoffstaetter96d}, one obtains
\begin{equation}
  \frac{\text{d}}{\text{d}\theta}\Phi =\kappa \; , \ \
  \frac{\text{d}}{\text{d}\theta}\vec S =
  \vec\Omega(\Phi)\times
  \vec S \; ,\ \
  \vec\Omega=\left(\begin{array}{c}\epsilon_\kappa\cos\Phi\\
                                   \epsilon_\kappa\sin\Phi\\
                                   \nu_0\end{array}\right)\; .
\label{eq:bmtsrm}
\end{equation}
Initial coordinates $\vec z_i$ are taken into final coordinates $\vec
z_f$ after one turn according to the relation
$\vec\Phi_f=\vec\Phi_i+2\pi\vec Q$ whence $\Phi_f=\Phi_i+2\pi\kappa$.
Now the orthogonal matrix $\underline T(\vec e,\varphi)$ is introduced to
describe a rotation around a unit vector $\vec e$ by an angle
$\varphi$.  Transforming the spin components of $\vec S$ into a
rotating frame using the relation $\vec S_R=\underline T(\vec
n_0,-\Phi)\cdot\vec S$, one obtains the simplified equation of spin
motion
\begin{equation}
\frac{\text{d}}{\text{d}\theta}\vec S_R  =
  \vec\Omega_R\times \vec S_R \; ,
\ \ \vec\Omega_R = \left(\begin{array}{c}
\epsilon_\kappa \\ 0 \\\delta \end{array}\right) \; , \ \ \delta=\nu_0-\kappa\; .
\end{equation}
If a spin field is oriented parallel to $\vec{\Omega}_R$ in this
frame, it does not change from turn to turn.  Therefore $\vec
n_R=\vec{\Omega}_R/|\vec{\Omega}_R|$ is an invariant spin field.  In the
original frame, this $\vec n$-axis is
\begin{equation}
  \vec n(\Phi) = \mathrm{sig}(\delta)\frac{1}{\Lambda}
  \left(\begin{array}{c}\epsilon_\kappa\cos\Phi\\
                     \epsilon_\kappa\sin\Phi\\
                     \delta\end{array}\right) \; ,\ \
    \Lambda = \sqrt{\delta^2 + \epsilon_\kappa^2} \; ,
\label{eq:srmn}
\end{equation}
where the `sign factor' $\mathrm{sig}(\delta)$ has been chosen so that on
the closed orbit ($\epsilon_\kappa=0$) the $\vec n$-axis $\vec
n(\Phi)$ coincides with $\vec n_0 = (0,0,1)^T$.  As required, $\vec n$
is both a solution of the T-BMT equation (\ref{eq:bmtsrm}),
$\frac{\text{d}}{\text{d}\theta}\vec n={\rm
  sig}(\delta)\frac{\kappa\epsilon_\kappa}{\Lambda}
(-\sin\Phi,\cos\Phi,0)^T=\vec\Omega\times\vec n$ and, as with any
function of phase space, a $2\pi$-periodic function of the angle variables
$\vec\Phi$ and of $\theta$.

This analytically solvable model can also be used to illustrate the
construction of a phase independent but amplitude-dependent spin tune
$\nu(\vec J)$.  Once an $\vec n$-axis has been obtained, one can
transform the components of $\vec S$ into a coordinate system
$[\vec{\tilde u}_1,\vec{\tilde u}_2, \vec n]$.  With the simple choice
\begin{eqnarray}
  \vec{\tilde u}_2(\Phi)
  &=&
  \frac{\vec n_0\times\vec n}{|\vec n_0\times\vec n|}
  =
  \mathrm{sig}(\delta)\left(\begin{array}{c}-\sin\Phi\\
                               \phantom{-}\cos\Phi\\
                              0\end{array}\right) \; ,\\
  \vec{\tilde u}_1(\Phi)
  &=&
  \frac{1}{\Lambda}\left(\begin{array}{c}\delta\cos\Phi\\
                                         \delta\sin\Phi\\
                     -\epsilon_\kappa\end{array}\right)\; ,
  \label{eq:srmu1t}
\end{eqnarray}
$\vec{\tilde u}_1$ is equal to $\vec{\tilde u}_2\times\vec n$ and the
basis vectors are clearly $2\pi$-periodic in $\vec\Phi$ and in
$\theta$ as required.  Since $\vec n$ and the basis vectors
$\vec{\tilde u}_1$ and $\vec{\tilde u}_2$ comprise an orthogonal
coordinate system for all $\theta$, and since $\vec n$ precesses
around $\vec\Omega$, one has
$\frac{\text{d}}{\text{d}\theta}\vec{\tilde
u}_2=(\vec\Omega-\tilde\nu\vec n)\times\vec{\tilde u}_2$ with the
rotation rate $\tilde\nu$ which can be computed by the relation
\begin{eqnarray}
\tilde\nu
&=&
(\frac{\text{d}}{\text{d}\theta} \vec{\tilde
u}_2-\vec\Omega\times\vec{\tilde u}_2) \cdot\tilde{\vec u}_1\nonumber\\
&=&
\mathrm{sig}(\delta)[
\left(\begin{array}{c}-\kappa\cos\Phi+\nu_0\cos\Phi\\
-\kappa\sin\Phi+\nu_0\sin\Phi\\ -\epsilon_\kappa\end{array}\right)
]\cdot\vec{\tilde u}_1\nonumber\\ &=& \mathrm{sig}(\delta)\Lambda\; .
\end{eqnarray}

\begin{figure}[htbp]
\begin{center}
\begin{minipage}[t]{\columnwidth}
\includegraphics[width=\columnwidth,bb=110 635 547 752,clip]
                {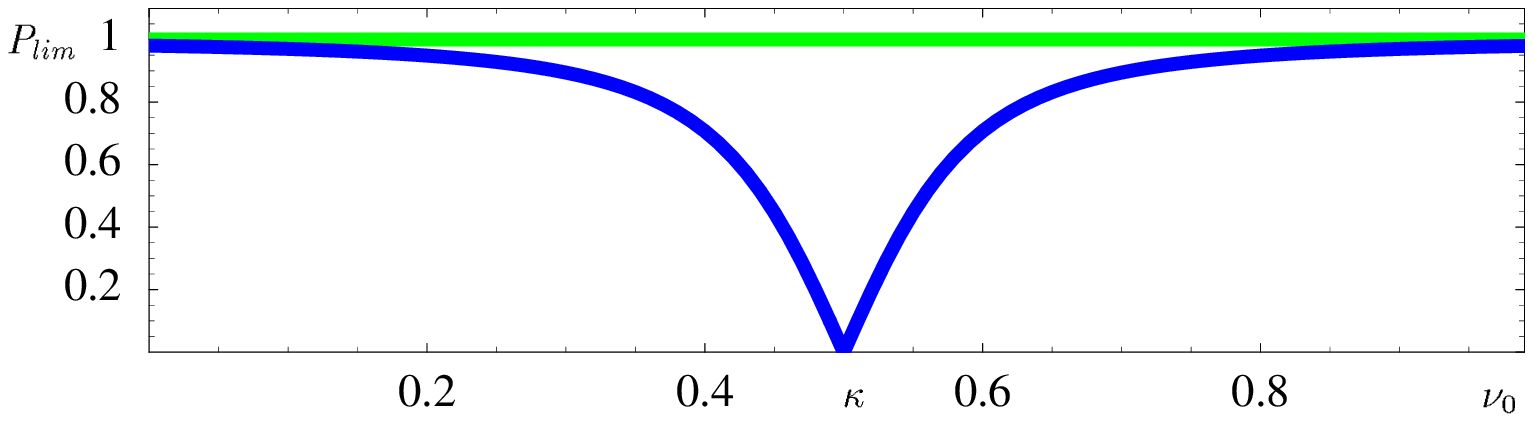}\\[3mm]
\includegraphics[width=\columnwidth,bb=109 635 547 752,clip]
                {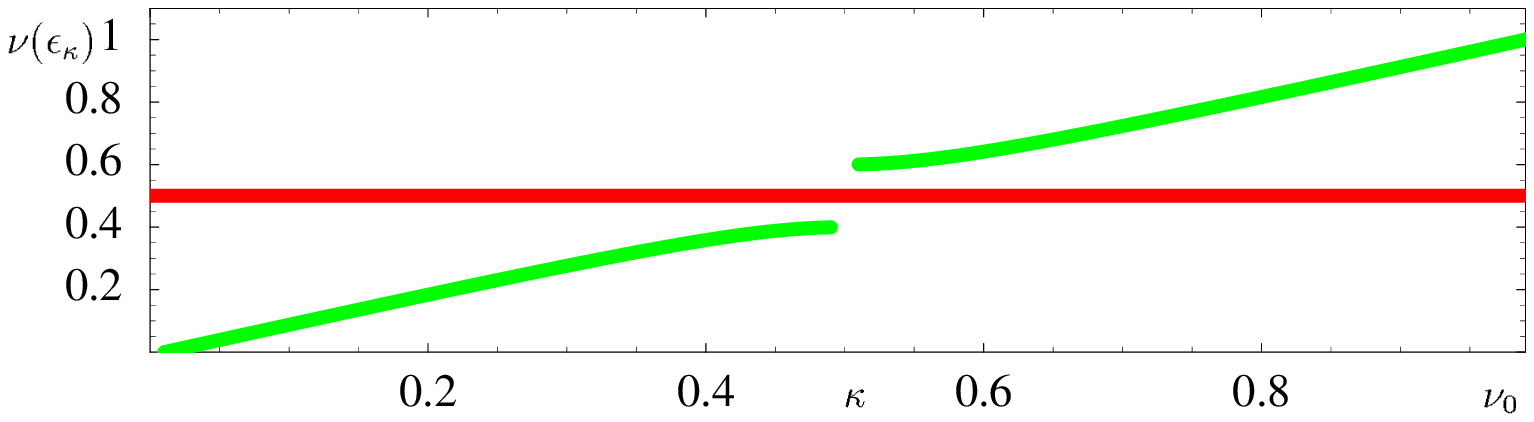}
\end{minipage}
\end{center}
\caption[$P_{\mathrm lim}$ and $\nu(\epsilon_\kappa)$ for the SRM]
 {$P_{\mathrm lim}$ and the amplitude-dependent spin tune
  $\nu(\epsilon_\kappa)$ for the SRM in the
  vicinity of $\nu_0=\kappa$, for $\kappa=0.5$ and
  $\epsilon_\kappa=0.1\;$
\label{fg:srmpnu}}
\end{figure}

In general, the so found rotation could depend on $\Phi$ and an
additional rotation of $\vec{\tilde u}_1$ and $\vec{\tilde u}_2$
around $\vec n$ can now be used to make $\tilde\nu$ independent of the
angle variables $\vec\Phi$ and to define the amplitude-dependent spin
tune.  Here however, $\tilde\nu$ is already independent of $\vec\Phi$
and it is therefore an amplitude dependent spin tune, and
$\epsilon_\kappa=|\vec\omega(\vec z)|$ characterizes the orbital
amplitude.  The freedom of rotating $\vec u_1$ and $\vec u_2$ around
$\vec n$ for each phase space point can be used to obtain a $\nu$
which reduces to $\nu_0$ on the closed orbit $(\epsilon_\kappa=0)$. We
let $\vec{\tilde u}_1$ and $\vec{\tilde u}_2$ rotate around $\vec n$
by $-\Phi$, to give the amplitude-dependent spin tune
\begin{equation}
  \nu = \mathrm{sig}(\delta)\Lambda + \kappa\; .
\label{eq:srmnu}
\end{equation}
The corresponding uniformly rotating basis vectors $\vec u_1$ and
$\vec u_2$ become
\begin{equation}
  \vec u_1=\vec{\tilde u}_1\cos\Phi-\vec{\tilde u}_2\sin\Phi\; ,\ \
  \vec u_2=\vec{\tilde u}_2\cos\Phi+\vec{\tilde u}_1\sin\Phi\; .
\label{eq:u12srm}
\end{equation}
On the closed orbit, the coordinate system now reduces to
\begin{equation}
\vec n\to\vec n_0\; ,\ \
\vec u_1\to \mathrm{sig}(\delta)\vec m \; ,\ \
\vec u_2\to \mathrm{sig}(\delta)\vec l \; ,\ \ \nu\to\nu_0\; .
\end{equation}
This model leads to the average polarization on the torus with
$\epsilon_\kappa(\vec J)$,
\begin{eqnarray}
P_{\mathrm lim}&=&|\langle\vec
  n(\vec z)\rangle| =
  \frac{|\delta|}{\sqrt{\delta^2+\epsilon_\kappa^2}}
=\sqrt{1-\left(\frac{\epsilon_\kappa}{\Delta}\right)^2}\; ,\\
\Delta&=&\nu-\kappa\; ,\ \delta=\nu_0-\kappa\; ,
\label{eq:srmplim}
\end{eqnarray}
where the distance of the amplitude-dependent spin tune $\nu$ from
the resonance has been denoted by $\Delta$, which is equivalent to ${\rm
sig}(\delta)\Lambda$.  In Fig.~\ref{fg:srmpnu}~(top) $P_{\mathrm lim}$ is
plotted versus $\nu_0$. It drops to 0 at $\nu_0=\kappa$ since
according to (\ref{eq:srmn}) the cone of vectors $\{\vec
n(\Phi)|\Phi\in[0,2\pi]\}$ opens up for small values of $|\delta|$.
This strong reduction of $P_{\mathrm lim}$ occurs when $\nu$ approaches
$\kappa$, i.e.~close to spin-orbit resonances.  According to
(\ref{eq:srmnu}) $\nu$ is never exactly equal to $\kappa$, but it
jumps by $2\epsilon_\kappa$ across the resonance condition
$\nu=\kappa$, which is shown in Fig.~\ref{fg:srmpnu}~(bottom).  This
jump of the spin tune could in principle be transformed away since the
sign of the spin tune depends on the sign of the rotation direction
$\vec n$. Here the sign of $\vec n$ in (\ref{eq:srmn}) has been fixed
by choosing $\vec n_0\cdot\vec n>0$, and the tune jump is therefore
essential.

Now we want to investigate the crossing of resonances, for the SRM,
and describe spin motion when a parameter $\tau$ of the system is
being slowly changed,
i.e.~$\frac{\text{d}}{\text{d}\theta}\tau=\alpha$.  In
particular this allows the study of an acceleration where $\nu_0$
crosses the frequency $\kappa$.  It is useful to describe the spin
motion in the coordinate system $[\vec u_1,\vec u_2,\vec n]$.  In
order to take account of the change of the basis vectors with the
parameter $\tau$, we use that for a vector with $|\vec u_i|=1$,
$\partial_\tau \vec u_i$ is perpendicular to $\vec u_i$ so that it can
be written as a rotation,
\begin{equation}
\frac{\partial}{\partial\tau}\vec n=
\vec\eta\times\vec n\ ,\ \
\frac{\partial}{\partial\tau}\vec u_1=
\vec\eta\times\vec u_1\ .
\end{equation}
The rotation vector $\vec\eta$ is then given by,
\begin{equation}
\vec\eta=\frac{1}{2}(
 \vec u_1\times\partial_\tau\vec u_1
+\vec u_2\times\partial_\tau\vec u_2
+\vec n  \times\partial_\tau\vec n)\; .
\label{eq:etacomp}
\end{equation}

Since $\nu_0=G\gamma$ in a flat ring, the acceleration process in the
SRM is usually described by a slowly changing
$\nu_0=\kappa+\tau$ with $\tau=\alpha\theta$ while assuming that
$\kappa$ and $\epsilon_\kappa$ do not change with energy.  This leads
to the following expressions for the variation of the basis vectors and
for $\vec\eta$:
\begin{eqnarray}
\partial_\tau\vec u_1
&=&
\mathrm{sig}(\delta)\frac{\epsilon_\kappa}{\Lambda^2}\vec n\cos\Phi\; ,\\
\partial_\tau\vec u_2
&=&
\mathrm{sig}(\delta)\frac{\epsilon_\kappa}{\Lambda^2}\vec n\sin\Phi\; ,\\
\partial_\tau\vec n
&=&
-\mathrm{sig}(\delta)\frac{\epsilon_\kappa}{\Lambda^2}\vec{\tilde u}_1\; ,\\
\vec\eta
&=&
\mathrm{sig}(\delta)\frac{\epsilon_\kappa}{\Lambda^2}
\frac{1}{2}(-\vec{\tilde u}_2-\vec u_2\cos\Phi+\vec u_1\sin\Phi)\nonumber\\
&=&
-\mathrm{sig}(\delta)\frac{\epsilon_\kappa}{\Lambda^2}\vec{\tilde u}_2\; .
\label{eq:srmeta}
\end{eqnarray}

In a general system, the equations of motion for the components of
$\vec S=\vec u_1s_1+\vec u_2s_2+\vec n J_S$ are described as
\begin{equation}
\frac{\text{d}}{\text{d}\theta}\left(\begin{array}{c}
s_1\\
s_2\\
J_S\\
\end{array}\right)
=
\left(\begin{array}{lllll}
\alpha(&\eta_3 s_2&-&\eta_2 J_S&)-\nu(\vec J,\tau)s_2\\
\alpha(&\eta_1 J_S&-&\eta_3 s_1&)+\nu(\vec J,\tau)s_1\\
\alpha(&\eta_2 s_1&-&\eta_1 s_2&)
\end{array}\right)\; .
\end{equation}
In complex notation with $\hat s=s_1+\text{i} s_2$,
$\eta=\eta_1+\text{i} \eta_2$, and $J_S=\sqrt{1-|\hat s|^2}$, this
gives
\begin{equation}
\frac{\text{d}}{\text{d}\theta}\hat s
=
i[\nu(\vec J,\tau)-\alpha\eta_3]\hat s
+
\text{i}\alpha\eta\sqrt{1-|\hat s|^2}\; .
\label{eq:shatode}
\end{equation}
For the SRM, the Eqs.~(\ref{eq:srmeta}) and (\ref{eq:u12srm})
lead to $\eta=-\text{i} \frac{\epsilon_\kappa}{\Lambda^2}
\text{e}^{\text{i} (\kappa\theta+\Phi_0)}$,
$\eta_3=0$, and
\begin{equation}
\frac{\text{d}}{\text{d}\theta}\hat s
=
i[\mathrm{sig}(\delta)\Lambda+\kappa]\hat s
+
\alpha\frac{\epsilon_\kappa}{\Lambda^2}\text{e}^{\text{i} (\kappa\theta+\Phi_0)}
\sqrt{1-|\hat s|^2}\; .
\label{eq:shatodesrm}
\end{equation}
Note again that the spin tune $\mathrm{sig}(\delta)\Lambda+\kappa$ in
this equation jumps by $2\epsilon_\kappa$ at $\nu_0=\kappa$.

We will now describe how this equation for the SRM leads to the
Froissart-Stora formula. After that, we will use the similarity of the
SRM in Eq.~(\ref{eq:shatodesrm}) and the equation for a general system
in Eq.~(\ref{eq:shatode}) to show how higher-order resonance strength
can be introduced and how they can be computed.

\section{The Froissart-Stora Formula}
\label{sc:froissart}

For the analytically solvable SRM the change of the adiabatic
invariant $J_S=\vec S\cdot\vec n$ can be computed explicitly.  When
the design-orbit spin tune changes during the acceleration process,
resonances will be encountered, where $\nu$ jumps from
$\kappa\pm\epsilon_\kappa$ to $\kappa\mp \epsilon_\kappa$ while the
spin is under the strong influence of an approximately resonant
Fourier contribution of $\omega$.  It is then found that for some
speeds of the spin tune change, parametrized by $\alpha$, a reduction
of polarization can occur is due to a generally irreversible reduction
of $J_S$ rather than a temporary decrease of $P_{\mathrm lim}$, and
which does not recover after the energy has increased and the
resonance is crossed.

To describe the reduction of polarization during resonance crossing,
(\ref{eq:shatodesrm}) can be used but the usual approach is to insert
a changing closed orbit spin tune $\nu_0$ into the equation of motion
(\ref{eq:bmtsrm}).  The method of solution depends on the form of the
function $\nu_0(\theta)$
\cite{froissart60,courant80,lee97a,schlesinger85}.  If the
closed-orbit spin tune changes like $\nu_0=\kappa+\alpha\theta$,
the equation of spin motion can be solved in terms of confluent
hypergeometric functions.  The equations for arbitrary initial
conditions are quite complicated but when at $\theta\to-\infty$ a
vertical spin $s_3(-\infty)=1$ is chosen as the initial condition then
the vertical component at $\theta\to+\infty$ is given by the well
known and regularly used Froissart-Stora formula \cite{froissart60},
\begin{equation}
s_3(\infty)=2\text{e}^{-\pi\frac{\epsilon_\kappa^2}{2\alpha}}-1\; .
\label{eq:froissart}
\end{equation}
In the case of a strong perturbation $\epsilon_\kappa$, or when the
acceleration is very slow, spins follow the change of $\vec n(\Phi)$.
The $\vec n$-axis in (\ref{eq:srmn}) has a discontinuity from $\vec
n_-=-\epsilon_\kappa(\cos\Phi,\sin\Phi,0)^T$ just below resonance to
$\vec n_+=-\vec n_-$ just above resonance.  Spins do not follow this
instantaneous change of sign, but they then follow $-\vec n$
adiabatically after the resonance has been crossed.  Therefore
$s_3(\infty)$ is close to $-1$ for a slow change of $\nu_0$.  When the
perturbation is weak or crossed very quickly, then spin motion is
hardly affected and $s_3(\infty)$ is close to 1 in
(\ref{eq:froissart}).  In intermediate cases, $|s_3|$ is reduced.  In
the first case the polarization is preserved but the spins are
reversed.  In the second case the polarization is preserved without
reversal.  In the third case the polarization is no longer vertical
but precesses around the vertical so that the time averaged
polarization is reduced.

\section{The Froissart-Stora Formula for Higher-Order Resonances}
\label{sc:nlfroiss}

As mentioned above, the Froissart-Stora formula in
Eq.(\ref{eq:froissart}) is regularly used to describe the reduction of
polarization due to vertical betatron motion during resonance crossing
in accelerators where the closed-orbit spin tune $\nu_0$ changes with
energy.  These descriptions were normally restricted to flat rings and
$\nu_0=G\gamma$. 

Since Siberian Snakes
\cite{derbenev76,derbenev78a,derbenev78b,krisch89a,luccio94} are
unavoidable for high-energy polarized beam acceleration, the
design-orbit spin tune is $\frac{1}{2}$ in most cases which will be
considered here and it does not change during acceleration.  Since the
orbital tunes are never chosen to be $\frac{1}{2}$, first-order
resonances with $\nu=j_0\pm Q_k$ are avoided and higher-order
resonances can become dominant.  But since the strength of such
resonances cannot be obtained as a Fourier coefficient of $\omega(\vec
z(\theta),\theta)$, a method for obtaining the strength of the
higher-order resonances is required in order to use the
Froissart-Stora formula when Siberian Snakes are in use.

HERA-p will require at least 4 Siberian Snakes
\cite{hoffstaetter00,hoffstaetter02b, hoffstaetter03,hoffstaetter04a}.
The snake angles $\varphi_j$ of these 4 snakes can be chosen quite
arbitrarily, except for the restriction
$\Delta\varphi=\varphi_4-\varphi_3+\varphi_2-\varphi_1=\frac{\pi}{2}$.
To illustrate crossing higher-order resonances a snake scheme for
HERA-p was chosen that has 4 Siberian Snakes with snake angles of
$\frac{\pi}{4}$, $0$, $\frac{\pi}{4}$ and $0$ in the South, East,
North and West straight section, respectively.

\begin{figure}[ht!]
\begin{center}
\begin{minipage}[t]{\columnwidth}
\includegraphics[width=\columnwidth,bb=127 534 548 753,clip]
                {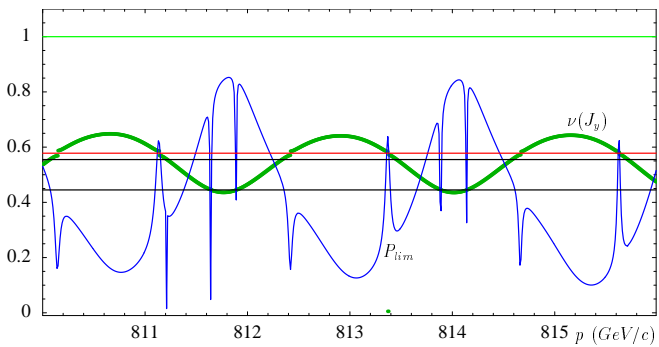}
\end{minipage}
\end{center}
\caption[$P_{\mathrm lim}$, the amplitude-dependent spin tune, and
  higher-order resonances for HERA-p]{$P_{\mathrm lim}$ (\emph{blue}) and
  $\nu(J_y)$ (\emph{green}) for particles with a $4.2\sigma$ vertical
  amplitude of $70\pi$~mm~mrad in HERA-p with and $Q_y=0.289\;$.
  Three resonance lines cross $\nu$ and at each crossing $P_{\mathrm lim}$
  exhibits a large variation and there are jumps in $\nu$, bottom:
  $\nu=5Q_y-1$, middle: $\nu=2-5Q_y$, and top
  $\nu=2Q_y$\label{fg:nlres}}
\end{figure}

In Fig.~\ref{fg:nlres} the amplitude-dependent spin tune (green) and
$P_{\mathrm lim}$ (blue) are plotted versus the reference momentum for a
vertical amplitude of $70\pi$~mm~mrad.  Many higher-order resonances
can be observed.  The curves for $P_{\mathrm lim}$ and $\nu(\vec J)$ were
computed with the non-perturbative algorithm SODOM~II \cite{yokoya99}
using the spin-orbit dynamics program SPRINT
\cite{man_sprint02,hoffstaetter96d}.  The $\vec n$-axis and also
$P_{\mathrm lim}$ are in general different at different azimuth $\theta_0$.
For this figure and for all following plots of $P_{\mathrm lim}$, the $\vec
n$-axis was observed at the interaction point of the ZEUS experiment
in the South of HERA.

While the design-orbit spin tune remains at $\frac{1}{2}$, the
amplitude-dependent spin tune $\nu(J_y)$ changes with energy and is in
resonance with $2Q_y$ at the second line (red) and with $5Q_y-1$ at
the bottom line at several energies.  In both cases a clear change of
$P_{\mathrm lim}$ can be observed.  The reduction of $P_{\mathrm lim}$
at some resonances is similar to the behavior for the single resonance
approximation shown in (\ref{eq:srmplim}) where $P_{\mathrm lim}$ is
reduced at those resonances. The drop of $P_{\mathrm lim}$ at
811.2~GeV/c is due to the $2-5Q_y$ resonance, which lies a little
below the $2Q_y$ line.  At all other energies where this resonance is
crossed, no influence on $P_{\mathrm lim}$ can be observed since the
corresponding fifth-order resonance strength is very small.  At some
second-order resonances, $P_{\mathrm lim}$ increases resonantly.
Presumably, two resonant effects are in constructive interference at
these energies.  Nonetheless, polarization can be reduced when these
resonance positions are crossed during acceleration since a sudden
increase of $P_{\mathrm lim}=$\mbox{$\langle\vec n\rangle$} is due to
a sudden change of $\vec n(\vec z)$ which might be too sudden for the
adiabatic invariance of $J_S=\vec S\cdot\vec n(\vec z)$ to be
maintained.  In addition one can see in Fig.~\ref{fg:nlres} that the
spin tune $\nu(J_y)$ has discontinuities at some of the resonances.

When spin motion in a ring is approximated by a single resonance with
$\kappa=j_0\pm Q_y$ and then Siberian Snakes are included in the ring,
it has often been noted that only odd-order resonances with
$\kappa=j_0+j_yQ_y$ appear, i.e.~$j_y$ is odd.  However it can be
shown by nonlinear normal form theory that this is a feature of any
ring with midplane symmetric spin-orbit motion and is not peculiar to
rings with Siberian Snakes \cite{vogt00}.  For rings without midplane
symmetry, resonances of even order can appear also. HERA-p has
non-flat regions, and rings with closed-orbit distortions in general
do not have midplane symmetric motion. Then, resonances with even
$j_y$ can also appear and be destructive.  In fact, the resonances
with $j_y=2$ are among the most destructive spin-orbit resonances in
HERA-p after Siberian Snakes are included. For the IUCF cooler ring
with a partial snake running, second-order resonances have been
observed experimentally \cite{alexeeva95a}.

When a parameter $\tau$ is being varied, the spin motion is described
in the coordinate system $[\vec u_1,\vec u_2,\vec n]$ by
Eq.~(\ref{eq:shatode}).  In the following we will demonstrate that
this equation has some characteristics of the equation of spin motion
(\ref{eq:shatodesrm}) of the SRM.  If the spin tune $\nu$ has a
discontinuity from $\nu_-$ to $\nu_+$ at some energy, then we define
the center frequency $\kappa^*=\frac{1}{2}(\nu_- + \nu_+)$.  To take
the jump of $\nu$ into account, we introduce
$\Lambda^*=|\nu-\kappa^*|$, which does not have a discontinuity and we
express the spin tune as $\nu={\rm
sig}(\nu-\kappa^*)\Lambda^*+\kappa^*$.

Since $\vec\eta$ is related to the basis vectors by
(\ref{eq:etacomp}), it is a $2\pi$-periodic function of $\vec\Phi$ and
$\theta$. The jump of $\nu$ across $\kappa^*$ can be produced by a
Fourier component of $\eta$ if there is a set of integers so that
$\vec j\cdot\vec Q+j_0=\kappa^*$.  This is the case in all instances
of spin tune jumps presented here.  Accordingly, one can analyze what
happens when the Fourier component $\eta_{\kappa^*} \text{e}^{\text{i}
(\kappa^*\theta+\Phi_0)}$ of $\eta$ dominates the motion of $\hat s$.
For that analysis, all other Fourier components of $\eta$ are ignored.
When $\alpha$ is small, spins which are initially almost
parallel to the $\vec n$-axis remain close to $\vec n$ so that $\hat
s$ is small and $\alpha\eta_3\hat s$ can therefore be ignored.
This leads to
\begin{equation}
\frac{\text{d}}{\text{d}\theta}\hat s
=
i(\mathrm{sig}(\nu-\kappa^*)\Lambda^*+\kappa^*)\hat s
+
I\alpha\eta_{\kappa^*} \text{e}^{\text{i} \kappa^*\theta+\Phi_0}
\sqrt{1-|\hat s|^2}\; .
\label{eq:shatapprox}
\end{equation}
Due to its similarity with (\ref{eq:shatodesrm}), this equation will
produce the observed spin tune jump by
$2\epsilon_{\kappa^*}=|\nu_+-\nu_-|$ if
$\eta_{\kappa^*}=\frac{\epsilon_{\kappa^*}}{\Lambda^{*2}}
=\frac{\epsilon_{\kappa^*}}{(\nu-\kappa^*)^2}$ in the vicinity of the
energy where the jump occurs.  Otherwise (\ref{eq:shatapprox}) would
not reproduce this jump.  One is then left with a relation which has
exactly the structure of the equation of motion (\ref{eq:shatodesrm})
for the SRM.  Therefore, the Froissart-Stora formula can be applied to
estimate how much polarization is lost when a polarized beam is
accelerated through the energy region where the spin tune jumps by
$2\epsilon_\kappa$.  In the following we will check whether, for some
higher-order resonances in HERA-p, all assumptions leading to the
approximation (\ref{eq:shatapprox}) are satisfied to the extent that
the Froissart-Stora formula describes the reduction of polarization
well.

The basis vectors $\vec n$, $\vec u_1$, and $\vec u_2$, and the
amplitude-dependent spin tune $\nu$ can in general only be computed by
computationally intensive methods.  The perturbing function $\eta$ is
then obtained from
\begin{eqnarray}
\eta
&=&
\vec\eta\cdot(\vec u_1+\text{i} \vec u_2)
=
\vec\eta\cdot(-\vec n\times\vec u_2+\text{i} \vec n\times\vec u_1)
\nonumber\\
&=&
(\vec\eta\times\vec n)\cdot(-\vec u_2+\text{i} \vec u_1)
=
i(\vec u_1+\text{i} \vec u_2)\cdot(\partial_\tau\vec n)\; ,
\end{eqnarray}
but the required differentiation is prone to numerical inaccuracies.
However, when $\vec n$ is computed by
perturbative normal form theory using
differential algebra (DA) \cite{balandin92}, the differentiation with
respect to $\tau$ can be performed automatically.  After $\eta$ is
computed, the Fourier integral over the complete ring would finally be
required in order to compute $\epsilon_\kappa$.

If (\ref{eq:shatode}) can be approximated well by a SRM, there is
however a different and much less cumbersome method for determining
the relevant resonance strength and the resonant frequency.
Observation of the amplitude-dependent spin tune $\nu(\vec J)$ allows
the determination of all parameters which are required to evaluate the
Froissart-Stora formula for higher-order resonances: The spin tune
jumps by $2\epsilon_\kappa$, the center of the jump is located at the
frequency $\kappa$ itself, and the rate of change of $\nu$ with
changing energy is used to determine the parameter $\alpha$ for
(\ref{eq:froissart}).  In the SRM this parameter is
$\frac{\nu_0-\kappa}{\theta}$ where $\nu_0$ is the frequency of spin
rotations when the resonance strength vanishes.  Here the
corresponding frequency, which would be observed if no perturbation
$\eta$ were present, is not directly computed.  But it can be
approximately inferred from the slope $\partial_\tau\nu$ at some
distance from the resonance.

\begin{figure}[ht!]
\begin{center}
\begin{tabular}{rr}
\begin{minipage}[t]{\columnwidth}
\includegraphics[width=0.49\columnwidth, bb=113 638 322 754, clip]
                {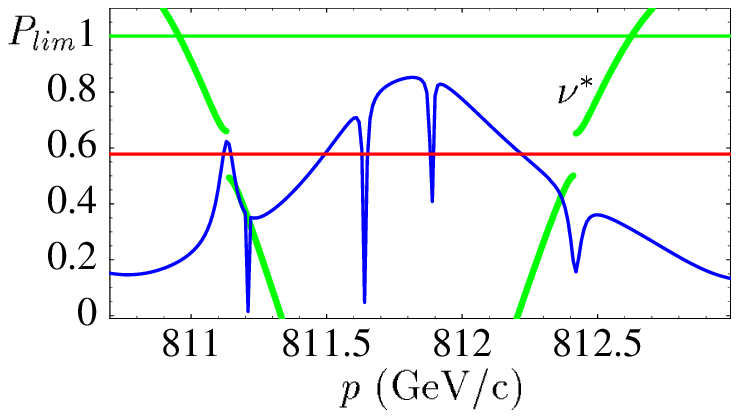}
\includegraphics[width=0.49\columnwidth, bb=124 644 322 760, clip]
                {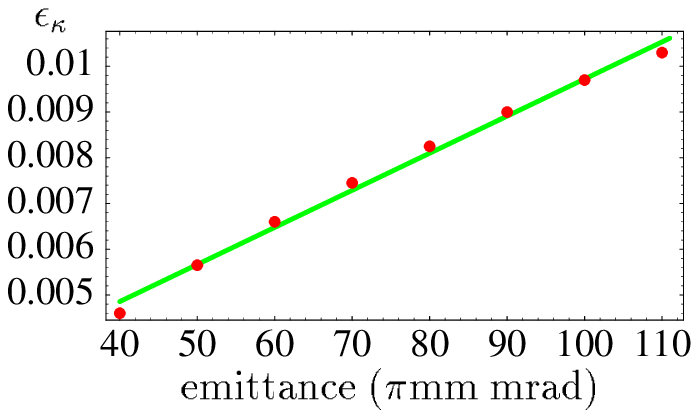}
\end{minipage}\\
\begin{minipage}[t]{\columnwidth}
\includegraphics[width=0.49\columnwidth, bb=118 650 324 758, clip]
                {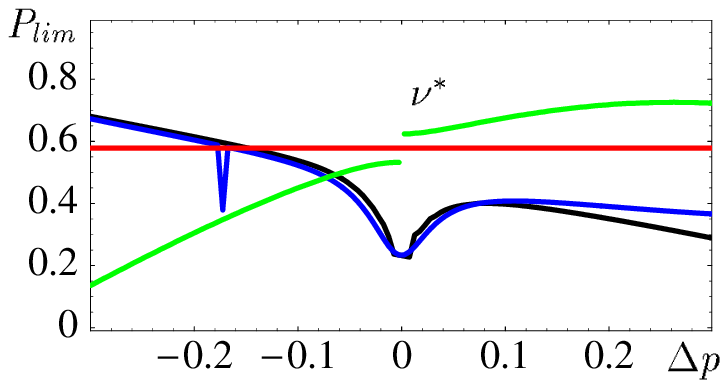}
\includegraphics[width=0.49\columnwidth, bb=118 650 324 758, clip]
                {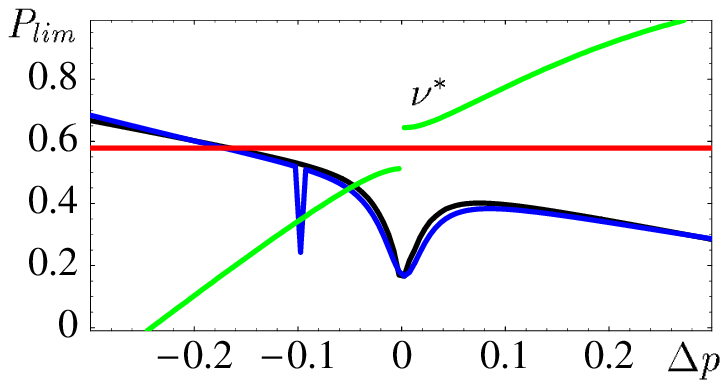}
\end{minipage}\\
\begin{minipage}[t]{\columnwidth}
\includegraphics[width=0.49\columnwidth, bb=118 650 324 758, clip]
                {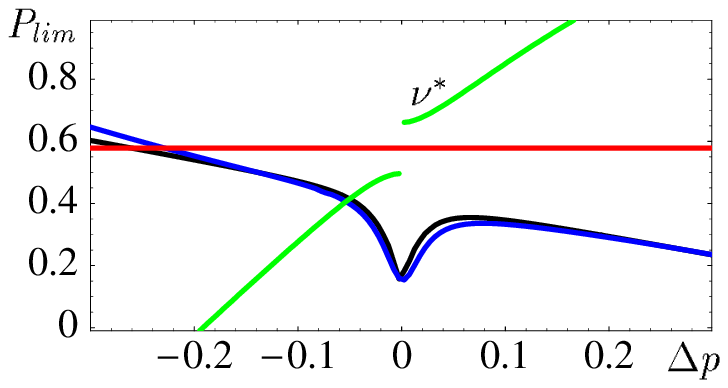}
\includegraphics[width=0.49\columnwidth, bb=118 650 324 758, clip]
                {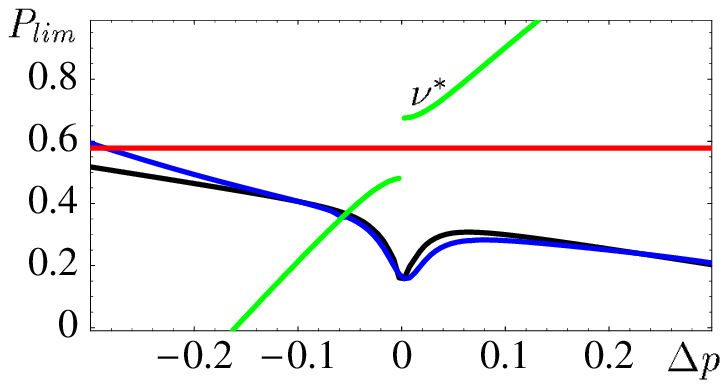}
\end{minipage}
\end{tabular}
\end{center}
\caption[$P_{\mathrm lim}$ and $\nu$ close to a resonance for different
vertical amplitudes] {{\bf Top left}: $P_{\mathrm lim}$ and $\nu$ in the
vicinity of the resonance at approximately 812.4~GeV/c for HERA-p. The
distance between $\nu$ and resonance has been magnified by 10,
$\nu^*=\kappa+10(\kappa-\nu)$.  {\bf Top right}: Proportionality
between tune jump $2\epsilon_k$ and the amplitude $2J_y$ of a vertical
emittance.  {\bf Middle and bottom}: Correlation between the width of
the actual drop of $P_{\mathrm lim}$ and the predictions of the single
resonance approximation using only the amplitude-dependent spin
tune. Vertical amplitudes of particles in HERA-p in units of
$\pi$~mm~mrad from top left to bottom right: 70, 40, 60.  80, and 100.
$\Delta p$: distance from the momentum at resonance in GeV/c
\label{fg:nlnup}}
\end{figure}

According to (\ref{eq:srmplim}), \mbox{$\langle\vec n\rangle$} is
  given by $P_{\mathrm lim}^{SRM}=
  \sqrt{1-(\frac{\epsilon_\kappa}{\nu-\kappa})^2}$ in the SRM.  To
  check whether the observed drop of $P_{\mathrm lim}$ indeed shows the
  characteristics of the SRM, the width of the resonance dip in
  $P_{\mathrm lim}^{SRM}$ was obtained from the amplitude-dependent spin tune
  alone and then compared to the width of the dip in the actual
  $P_{\mathrm lim}$ of the system.  This analysis was done for HERA-p's
  resonance at approximately 812.4~GeV/c and the results are shown in
  Fig.~\ref{fg:nlnup}. The top left plot shows the dependence of
  $P_{\mathrm lim}$ and $\nu$ on the reference momentum for a vertical
  amplitude of $70\pi$~mm~mrad which, with HERA-p's current one sigma
  emittance of 4$\pi$~mm~mrad, corresponds to the amplitude of a
  $4.2\sigma$ vertical emittance.  The momentum range is as in
  Fig.~\ref{fg:nlres}.  The low $P_{\mathrm lim}$ shows that many perturbing
  effects interfere in this region.  In units of $\pi$~mm~mrad, the
  vertical amplitude of the particles in the top left graph is 70, in
  the middle graphs it is 40 and 60, and in the bottom graphs 80 and
  100.  The horizontal scale displays the distance $\Delta p$ in
  $GeV/c$ from the momentum at the resonance.

In the 4 bottom graphs, $P_{\mathrm lim}$ and $P_{\mathrm lim}^{SRM}$ are plotted for
different orbital amplitudes, and the different resonance strengths
are obtained from the jump in $\nu(J_y)$.  Only information about
$\nu$ was used to compute $P_{\mathrm lim}^{SRM}$.  To allow better
comparison, a linear change of $P_{\mathrm lim}^{SRM}$ with momentum was added
as a background curve and the height of the dip was scaled to fit the
actual $P_{\mathrm lim}$.  The width however was not changed.  The distance
between spin tune and resonance has been magnified by 10,
$\nu^*=\kappa+10(\kappa-\nu)$ in these graphs.  The tune jump is
symmetric around the resonance line $\nu=2Q_y$, showing that a
second-order resonance is excited.

As shown in Fig.~\ref{fg:nlnup}~(top right) the tune jump scales
approximately linearly with the orbital action variable $J_y$.  This
is consistent with the crossing of a second-order resonance, since a
frequency of $2Q_y$ can be produced by monomials of
$\sqrt{J_y}\text{e}^{\pm\text{i} Q_y\theta}$ with order larger or
equal to 2.  This linear scaling is not exact for two reasons: (1) The
jump does not reduce to 0 at $J_y=0$ but already at some finite
amplitude at which $\nu(J_y)$ does not cross the resonance line. (2)
When the amplitude is changed, the momentum at which the resonance
occurs changes, and the resonance strength is in general different at
different energies.  Deviations from a linear dependence should
therefore be expected.  $P_{\mathrm lim}$ is already very low away from the
resonance at $\nu=2Q_y$, indicating that other strong perturbations
distort the invariant spin field and can interfere with the resonance
harmonic.

Thus we conclude that the resonance width computed in terms of the
tune jump $2\epsilon_\kappa$ agrees surprisingly well with the actual
drop in $P_{\mathrm lim}$.

Since the higher-order resonances analyzed here show the established and
characteristic relation between tune jump and reduction of $P_{\mathrm lim}$,
the applicability of the Froissart-Stora formula will now be tested.

\begin{figure}[ht!]
\begin{center}
\begin{minipage}[t]{\columnwidth}
\includegraphics[width=0.49\columnwidth,bb=113 347 336 755]
                {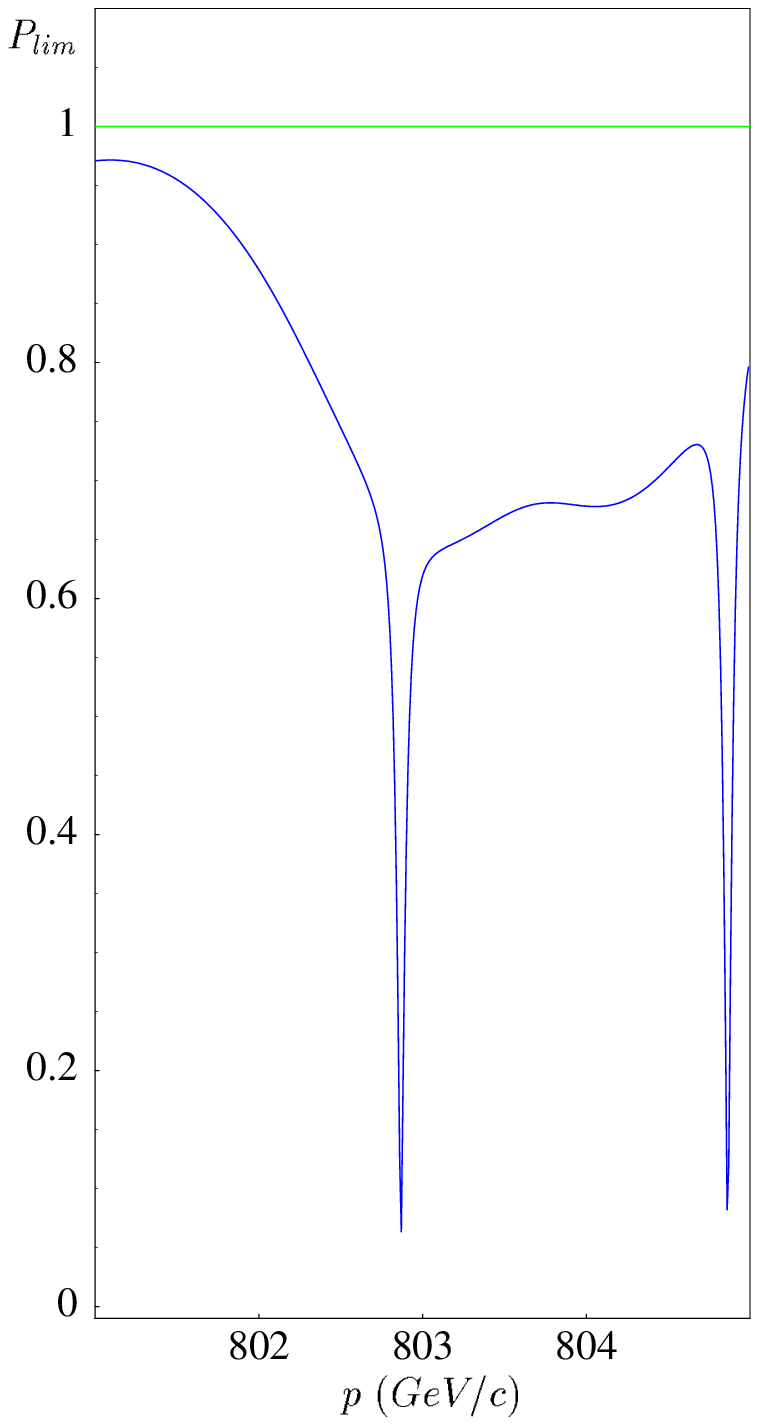}
\includegraphics[width=0.49\columnwidth,bb=121 358 336 762]
                {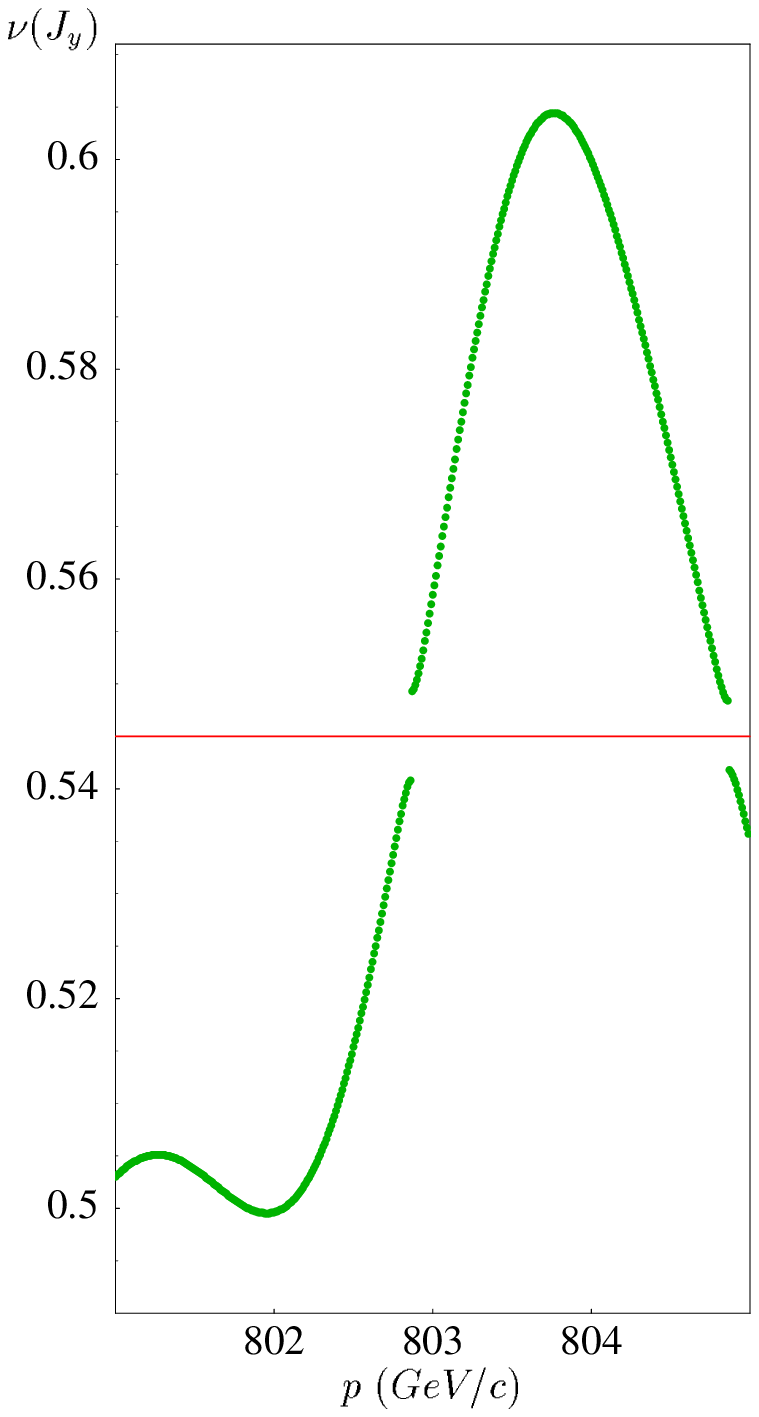}
\includegraphics[width=\columnwidth,bb=110 523 547 754,clip]
                {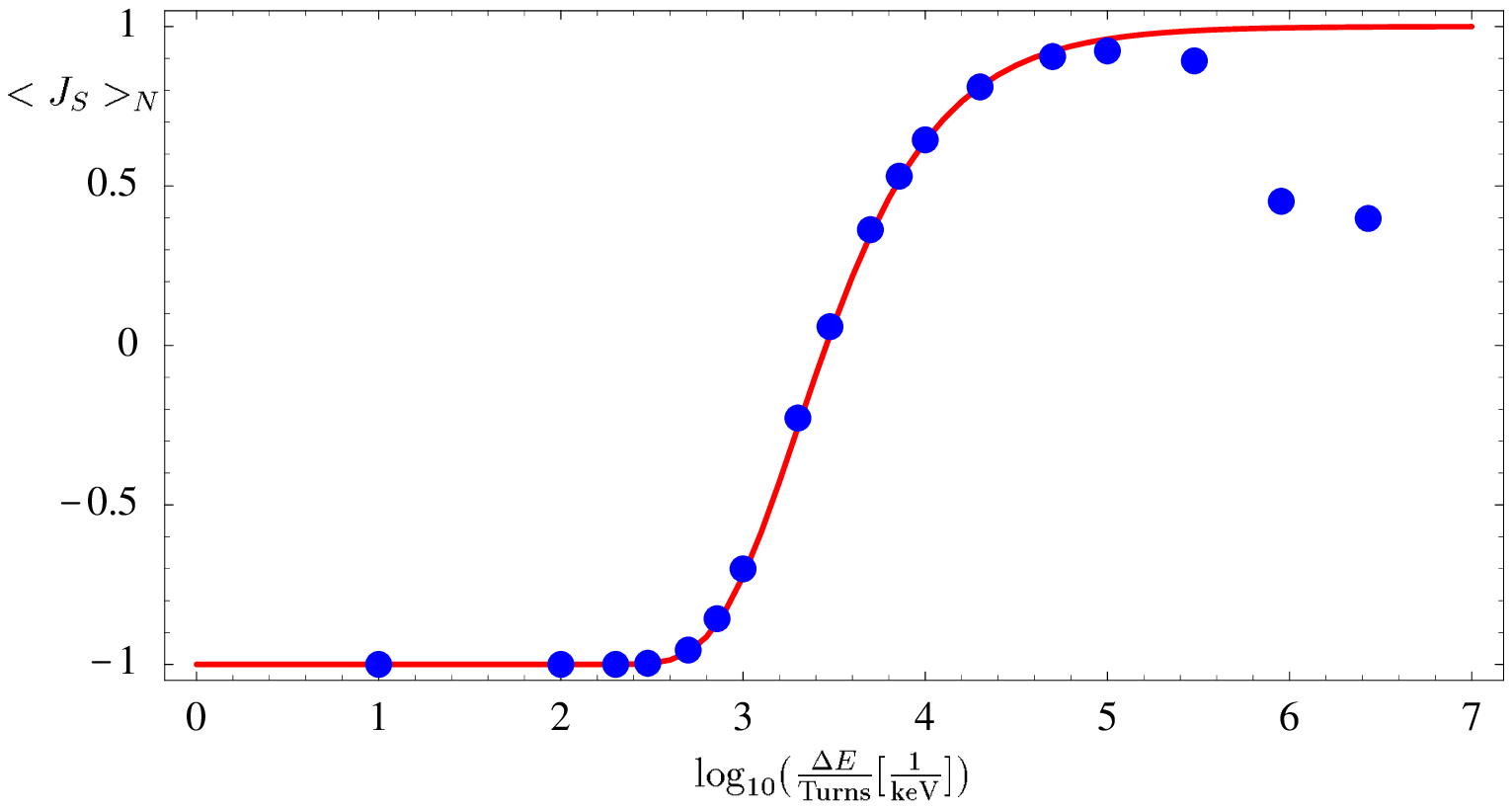}
\end{minipage}
\end{center}
\caption[Predictability of polarization loss at higher-order
resonances for HERA-p] {{\bf Top}: $P_{\mathrm lim}$ and $\nu$ for a
second-order resonance of HERA-p with $Q_y=0.2725$ and a
$0.75\sigma$ vertical amplitude of $2.25\pi$~mm~mrad.  {\bf Bottom}:
\mbox{$\langle J_S\rangle_N$} after acceleration from 801~GeV/c to
804~GeV/c with different acceleration rates (\emph{blue points}) and
the prediction of the Froissart-Stora formula (\emph{red curve}) using
parameters $\epsilon_{2Q_y}$ and $\alpha$ obtained from $\nu$
\label{fg:hp96froissart}}
\end{figure}

In Fig.~\ref{fg:hp96froissart}~(top) $P_{\mathrm lim}$ and $\nu$ are shown for
HERA-p.  $P_{\mathrm lim}$ is reduced at two resonances with $\nu=2Q_y$.  The
vertical tune had been chosen as $Q_y=0.2725$ so that these resonances
are crossed already for the small $0.75\sigma$ vertical amplitude of
$2.25\pi$~mm~mrad.  At this small amplitude $P_{\mathrm lim}$ is reasonably
large.

The spins of a set of particles were set parallel to the invariant
spin field $\vec n(\vec z)$ so that all had $J_S=1$ at the momentum of
801~GeV/c.  The $\vec n$-axis had been computed by stroboscopic
averaging \cite{hoffstaetter96d}.  Due to the rather large $P_{\mathrm lim}$
at that energy the initial polarization was approximately 97\%.
Starting with this spin configuration, the beam was accelerated to
804~GeV/c at various rates.  The average \mbox{$\langle J_S\rangle_N$}
over the tracked particles is plotted versus acceleration rate in
Fig.~\ref{fg:hp96froissart}~(bottom) together with the prediction of
the Froissart-Stora formula.  The average \mbox{$\langle
J_S\rangle_N$} describes the degree of beam polarization which could
be recovered due to the adiabatic invariance of $J_S$ when moving into
an energy regime where $\vec n(\vec z)$ is close to parallel to the
vertical.

The resonance strength $\epsilon_{2Q_y}$ has been determined from the
tune jump.  The parameter $\alpha$ is proportional to the energy
increase per turn $d_E$ and is determined from the tune slope
$\frac{\Delta\nu}{\Delta E}$ in Fig.~\ref{fg:hp96froissart}~(top
right) by the relation $\alpha=\frac{1}{2\pi}\frac{\Delta\nu}{\Delta
  E}d_E$.

The polarization obtained by accelerating particles through the
second-order resonance agrees remarkably well with the Froissart-Stora
formula.  For the slow acceleration of about $50$~keV per turn in
HERA-p, the polarization would be completely reversed on the $0.75$
sigma invariant torus.  This would lead to a net reduction of beam
polarization, since the spins in the center of the beam are not
reversed.

This result on the applicability of (\ref{eq:froissart}) for
the resonance strength and $\alpha$ obtained from the amplitude
dependent spin tune is so important for detailed analysis of the
acceleration process that it will be checked in another case.
In the next example, the same lattice is used,
the tune was adjusted to a realistic value of $Q_y=0.289$ and a
$4.2\sigma$ vertical amplitude of $70\pi$~mm~mrad was chosen.  At this
large amplitude, the second and fifth-order resonances already shown
in Fig.~\ref{fg:nlres} are observed.  Particles were then
accelerated from 812.2~GeV/c to 812.6~GeV/c with different
acceleration rates.  Note that the initial condition has a vertical
polarization of only 60\%.  Nevertheless this state of polarization
corresponds to a completely polarized beam, and 100\% polarization can
potentially be recovered by changing the energy adiabatically into a
region where $\vec n(\vec z)$ is tightly bundled.  These studies
emphasize again the importance of choosing $\vec n(\vec z)$ as the
initial spin direction.  For example if the spins were initially
polarized vertically, they would rotate around $\vec n(\vec z)$ and that
would lead to a fluctuating polarization, even without a resonance and
it would not be possible to establish a Froissart-Stora formula for
higher-order resonances.

\begin{figure}[ht!]
\begin{center}
\begin{minipage}[t]{\columnwidth}
\includegraphics[width=0.49\columnwidth, bb=113 347 336 755]
                {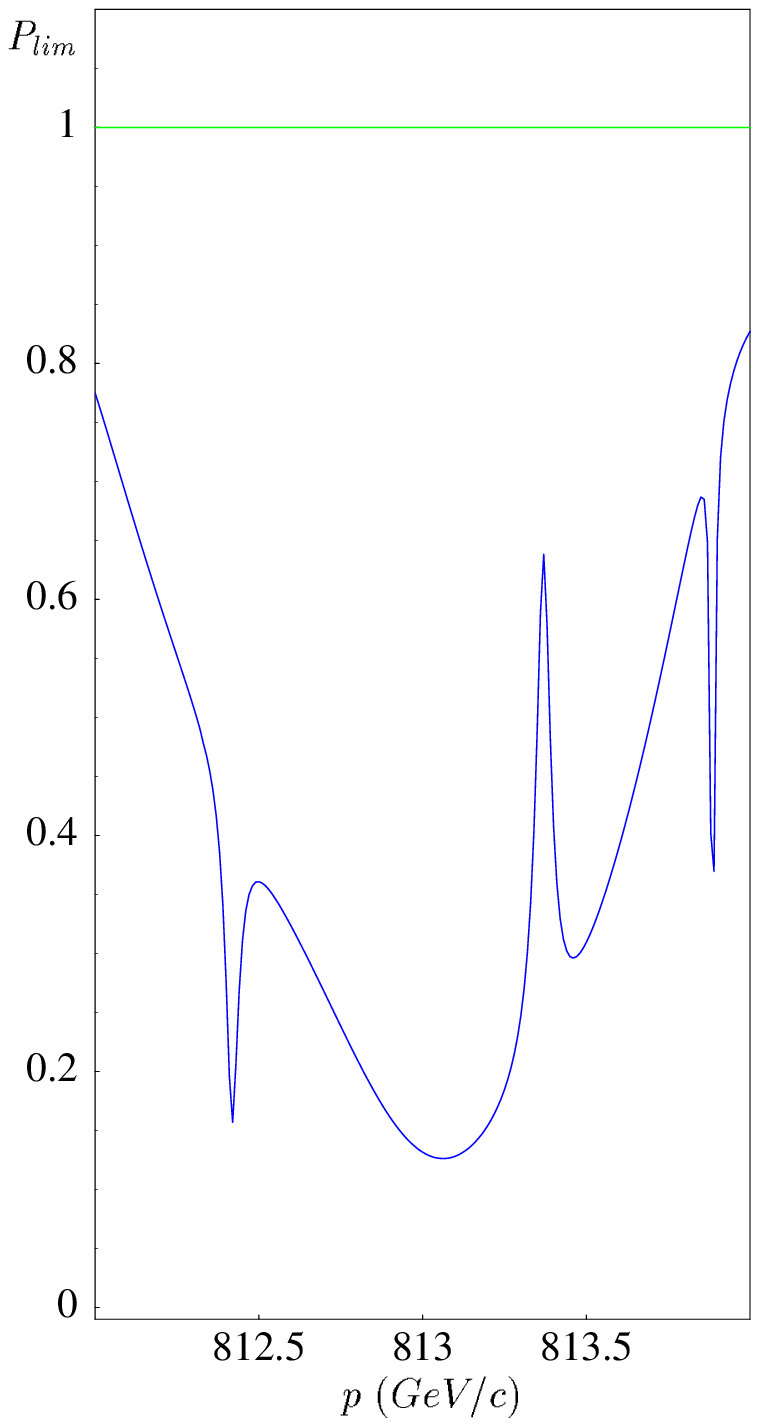}
\includegraphics[width=0.49\columnwidth, bb=121 358 336 755]
                {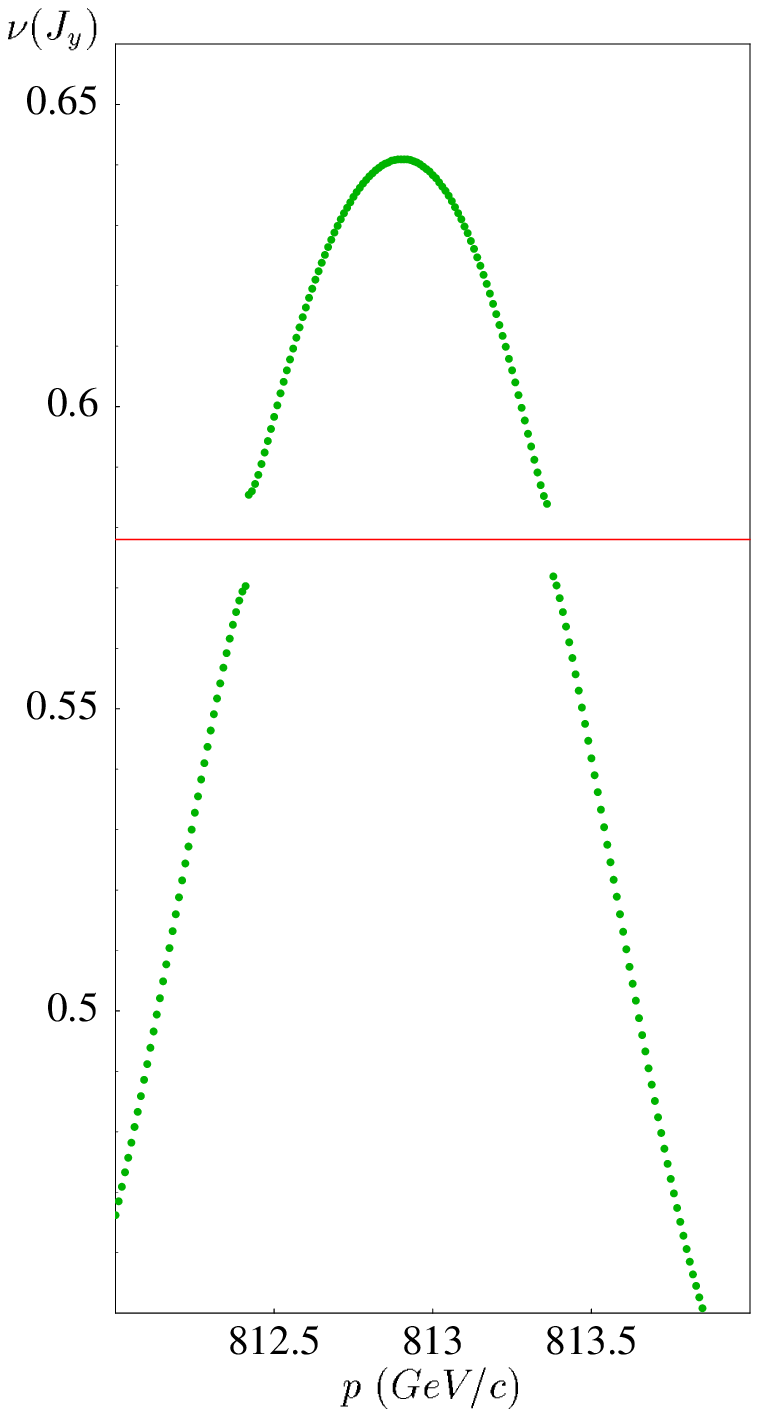}
\includegraphics[width=\columnwidth,bb=110 519 547 754, clip]
                {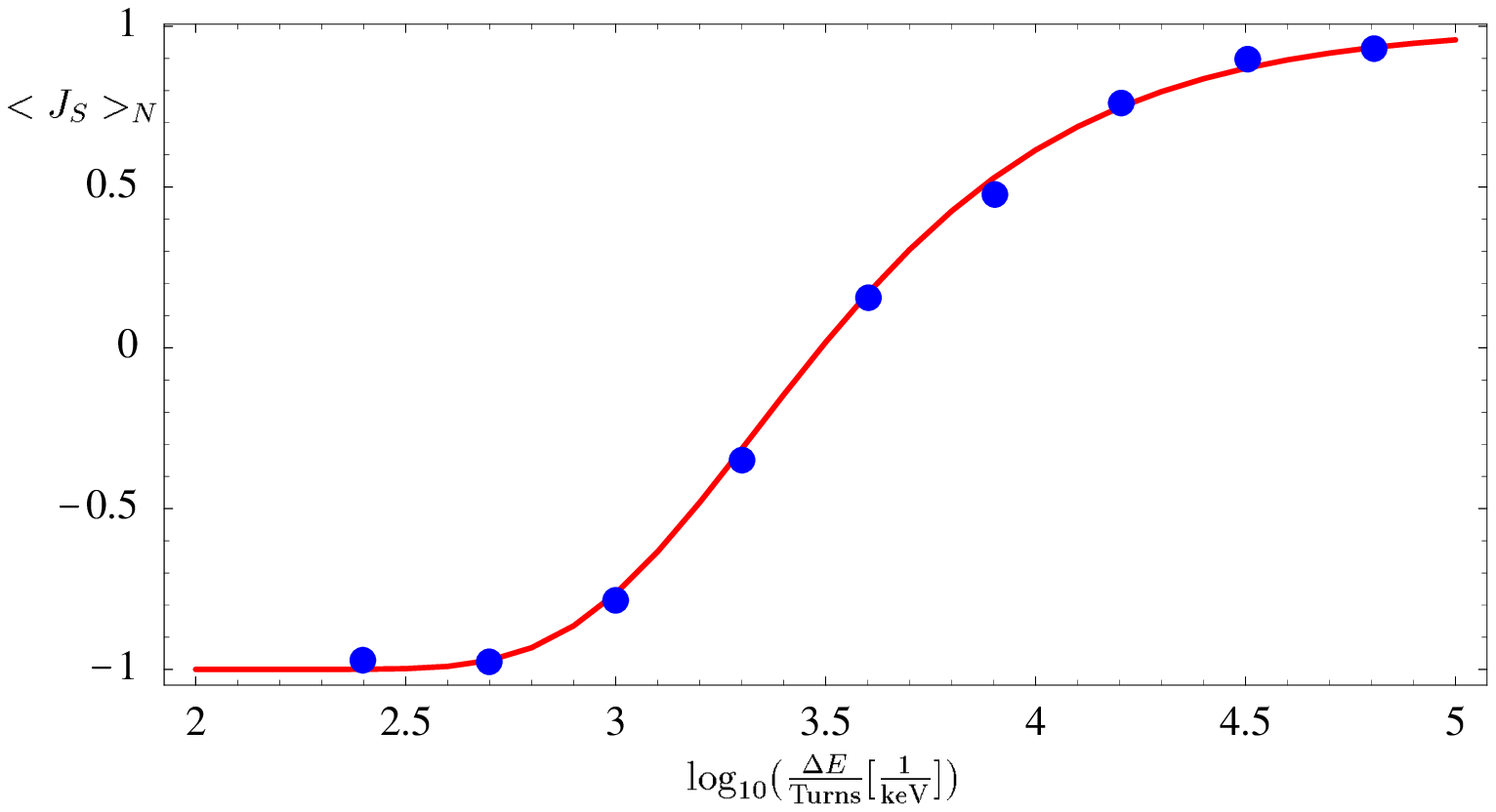}
\end{minipage}
\end{center}
\caption[Predictability of polarization loss at higher-order
resonances for HERA-p] {{\bf Top}: $P_{\mathrm lim}$ and $\nu$ for a
second-order resonance of HERA-p with realistic tune of $Q_y=0.289$
and a large $4.2\sigma$ vertical amplitude of $70\pi$~mm~mrad.  {\bf
Bottom}: \mbox{$\langle J_S\rangle_N$} after acceleration from
812.2~GeV/c to 812.6~GeV/c with different acceleration rates
(\emph{blue points}) and the prediction of the Froissart-Stora formula
(\emph{red curve}) using parameters $\epsilon_{2Q_y}$ and
$\alpha$ obtained from $\nu$
\label{fg:hpupgfroissart}}
\end{figure}

As shown in Fig.~\ref{fg:hpupgfroissart}, $P_{\mathrm lim}$ is as low as 0.11
in the center of the displayed region.  Obviously other strong effects
beyond the second-order resonance are present and overlap with it.
The bottom figure shows \mbox{$\langle J_S\rangle_N$} after the
acceleration.  The fact that \mbox{$\langle J_S\rangle_N$} is again
described very well by the Froissart-Stora formula
(\ref{eq:froissart}) is an impressive confirmation of the conjecture.

The two data points at largest acceleration speed in
Fig.~\ref{fg:hp96froissart}~(bottom) are lower than predicted by the
Froissart-Stora formula.  A possible explanation is the following: at
very large acceleration speeds the resonance region is crossed so
quickly that the spin motion is hardly disturbed.  But when the $\vec
n$-axis $\vec n_-$ before the resonance region is not parallel to the
$\vec n$-axis $\vec n_+$ after the resonance region, then the spins
which initially had $J_S=1$ will approximately have $J_S=\vec
n_-\cdot\vec n_+$ after the resonance region is crossed, which is
smaller than the Froissart-Stora prediction, which approaches 1 for
large acceleration speeds.

Here the parameter $\tau$ was the slowly changing momentum.  This
generalized way of using the Froissart-Stora formula can however also
be used when other system parameters change.  An example can be found
in \cite{hoffstaetter98d}, where the particle's phase space amplitude
is changed artificially slowly in order to compute the invariant spin
field at various orbital amplitudes.  In \cite{vogt00} an example is
displayed where the Froissart-Stora formula is successfully applied to
a resonance which is encountered because of a slow variation of $Q_y$.

\section{The Choice of Orbital Tunes}
\label{sc:tunes}

When the amplitude-dependent spin tune $\nu(\vec J)$ of particles
with the amplitude $\vec J$ crosses a resonance, for example during
acceleration, the beam polarization is usually reduced.
\begin{figure}[ht!]
\begin{center}
\begin{minipage}[t]{\columnwidth}
\includegraphics[width=0.49\columnwidth,bb=124 539 335 761,clip]
                {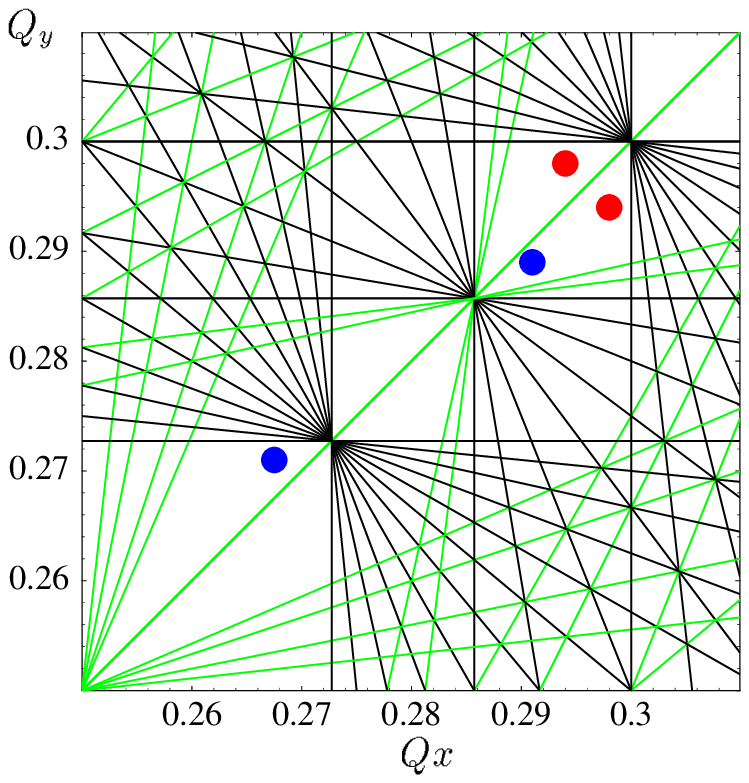}
\hfill
\includegraphics[width=0.49\columnwidth,bb=124 542 335 761,clip]
                {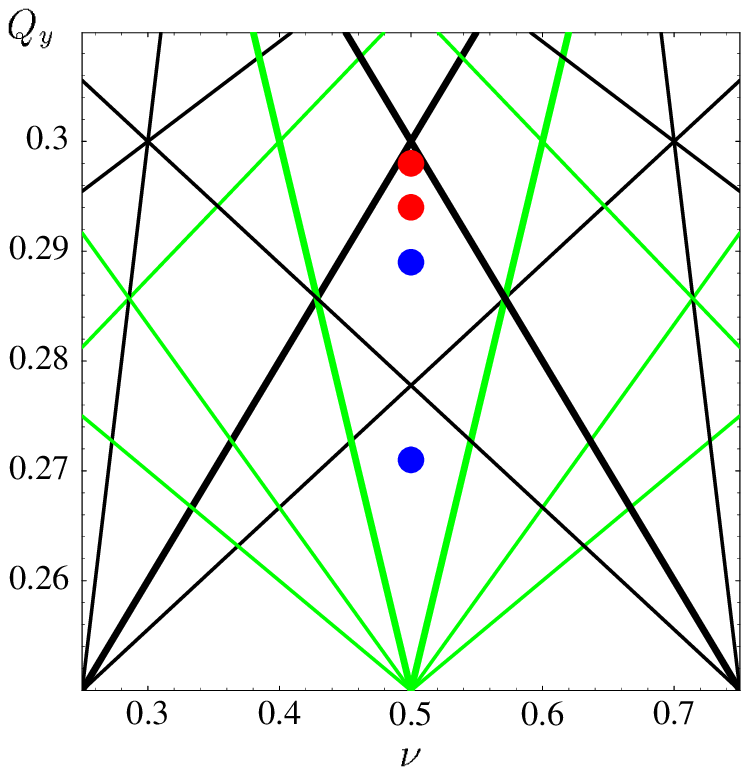}
\end{minipage}
\end{center}
\caption[Tune diagrams for $(Q_x,Q_y)$ and for $(\nu,Q_y)$]
 {{\bf Left}: the current orbit tunes ($Q_x=0.294$,$Q_y=0.298$)
  or ($Q_x=0.298$,$Q_y=0.294$) (\emph{red}) and the new orbit tunes for
  polarized proton operation ($Q_x=0.291$,$Q_y=0.289$) or
  ($Q_x=0.2675$, $Q_y=0.271$) (\emph{blue}) in the $x$-$y$ resonance diagram.
  All resonances up to order 11 are shown.  Difference resonances are
  indicated in green.  {\bf Right}: The current vertical tunes
  (\emph{red}) and the
  new vertical tunes (\emph{blue}) in the spin-orbit resonance diagram.  The
  odd spin-orbit resonances (\emph{black}) and the
  even spin-orbit resonances (\emph{green})
  are shown up to order 10 in the vicinity of closed-orbit spin tune
  $\nu_0=\frac{1}{2}$. For HERA-p, the resonances of second
  order (\emph{fat green}) and of fifth order (\emph{fat black}) are most
  destructive
\label{eq:tunes}}
\end{figure}
It is therefore important to find suitable orbital tunes so that
low-order spin-orbit resonances are far away from the operating point.
In particular, when Siberian Snakes are used to maintain a closed
orbit spin tune of $\frac{1}{2}$, it is important that these Snakes
are optimized so that higher-order resonances do not lead to large
deviations of the amplitude dependent spin tune from this value. Such
optimal choices of snakes are discussed in \cite{hoffstaetter04a}. The
dominant effects are due to radial fields on vertical betatron
trajectories.  Thus Fig.~\ref{eq:tunes}~(right) shows the resonance
lines $\nu=j_0+j Q_y$ up to order 10 in the $\nu$-$Q_y$ plane.  If the
spin tune on the closed orbit is fixed to $\nu_0=\frac{1}{2}$ by
Siberian Snakes the orbital tune can be chosen to avoid resonance
lines.  However, the dynamic aperture of proton motion should not be
reduced and the tunes have to be far away from low order orbital
resonances.  Figure (\ref{eq:tunes})~(left) shows the $Q_x$-$Q_y$ tune
diagram with resonance lines up to order 11.  The operating point has
to stay away from these resonance lines.

The established tunes of HERA-p operation $Q_x=0.294$, $Q_y=0.298$ or
$Q_x=0.298$, $Q_y=0.294$ (red points) would be unfortunate choices due
to their closeness to the resonance $\nu=j_0\pm 5Q_y$.  For HERA-p
with Siberian Snakes, several simulations have shown that the
resonances of second order and of fifth order are most
destructive. This is supported by Fig.~\ref{fg:nlres}.  Therefore two
new tunes (blue points) are suggested which have an optimal distance
from low-order spin-orbit resonances. It has been tested
experimentally that HERA-p could operate at these tunes.

\section*{Acknowledgment}
\begin{minipage}{\textwidth}
   Desmond Barber's careful reading and improving of the 
   manuscript are thankfully acknowledged.
\end{minipage}

\end{document}